\documentclass[preprint,showpacs,preprintnumbers,amsmath,amssymb]{revtex4}

\usepackage{bm}

\newcommand{\la}{\lambda}
\newcommand{\om}{\omega}

\newcommand{\La}{\Lambda}

\newcommand{\ra}{\rightarrow}

\newcommand{\qt}{\tilde{q}}
\newcommand{\Qt}{\tilde{Q}}

\newcommand{\phih}{\hat{\phi}}
\newcommand{\xh}{\hat{x}}
\newcommand{\yh}{\hat{y}}

\newcommand{\N}{\mathcal{N}}

\newcommand{\half}{\frac{1}{2}}

\begin{document}

\preprint{NORDITA-2003-90 HE}

\title{Gaugino and meson condensates in ${\cal N}=1$ SQCD from Seiberg-Witten curves}
\author{Michael Chesterman}
 \email{Michael.Chesterman@kau.se}
\affiliation{Department of physics, Karlstad University, S-651 88
Karlstad, Sweden}
\affiliation{Nordita, Blegdamsvej 17, Copenhagen,
DK-2100, Denmark.}

\date{November 2003}

\begin{abstract}
We calculate gaugino and meson condensates in $\N=1$ SQCD theory
with $SU(N_c)$ gauge group and $N_f < 2N_c$ matter flavours, by
deforming the pure $\N =2$ Super-Yang-Mills plus fundamental
matter action with a mass term for the adjoint scalar superfield.
This follows similar recent work by Konishi and Ricco for the case
without fundamental matter.
\end{abstract}

\maketitle

\section{Introduction} \label{sec:introduction}
The gaugino and meson vacuum condensates of SQCD are fundamental
features of the theory, which are associated with the presence of
a mass gap, confinement and instantons. Over the years, various
methods have emerged for calculating the values of these
condensates.

An indirect but simple way is to calculate the minima of one of
two related non-perturbative quantum superpotentials: the
Veneziano-Yankielovicz-Taylor (VYT) superpotential
\cite{Taylor:1983bp}, which was conjectured on the basis of
reproducing the correct $U(1)$ anomalies of the theory, or the
Affleck-Dine-Seiberg (ADS) superpotential \cite{Affleck:1984mk},
which was deduced using holomorphy arguments, and the $U(1)_A$
selection rule. For the ADS superpotential, which only yields the
meson condensate, one can obtain the gaugino condensate by using a
suitable Konishi anomaly equation. Note that both of these
superpotentials only exist when the number of flavours is less
than the number of colours $N_f<N_c$. For the region $N_f\geq
N_c$, alternative techniques for calculating the condensates are
needed.

There are also direct methods, as described in
\cite{Konishi:2003ts}. The so-called strong coupling instanton
(SCI) and weak coupling instanton (WCI) one-loop calculations
\cite{Dorey:2002ik}, though developed for calculating the gaugino
condensate in pure $\N=1$ SYM, can also be applied to SQCD by
taking into account the differing one-loop beta function. The
meson condensate can then be obtained from a Konishi anomaly
equation. The results of the two methods disagree by a factor of
4/5, though it is widely believed that the SCI calculation is not
reliable \cite{Dorey:2002ik} and that we should believe the WCI
calculation, which also agrees with the results from the VYT and
ADS superpotentials. The main problem with the SCI approach seems
to be that in the strong coupling confining phase of the theory,
semi-classical instanton calculations are not valid. In another
interesting technique \cite{Davies:1999uw} which could be
generalized to the case of fundamental matter, space-time is
compactified on $R^3\times S^1$, in order that the semi-classical
instanton calculation is valid.

In our approach, we use the intimate relationship between $\N=1$
and $\N=2$ SYM, extending Konishi and Ricco's calculation
\cite{Konishi:2003ts} to the case of fundamental matter with
$SU(N_c)$ gauge group and $N_f<2N_c$ flavours. In turn, their work
was a generalization of an article by Finnell and Pouliot
\cite{Finnell:1995dr} for $SU(2)$. It may come as no surprise that
our results agree with the prediction of the VYT and ADS
superpotentials, and the WCI calculation. Nevertheless, it is
satisfying, to have confirmation from another independent source.
Our method is also valid for a wider range of flavours than for
the two superpotentials.





\section{The Calculation}

As previously mentioned, starting with $\N=2$ super Yang-Mills
with $SU(N_c)$ gauge group and $N_f< 2N_c$ flavours, we add a mass
term $\mu \text{Tr } \phi^2$ for the $\N=1$ adjoint scalar field
$\phi$. In terms of the resultant $\N=1$ theory, the classical
superpotential is given by
\begin{equation}\label{eq:W_tree}
W_{tree} = \mu \text{Tr }\phi^2 + \sqrt{2} \text{Tr }
\sum_{f=1}^{N_f}{\Qt_f \phi Q^f} + \text{Tr }\sum_{f=1}^{N_f}
\Qt_f m_f Q^f,
\end{equation}
where $Q^f$ and $\Qt_f$ are quark superfields in the fundamental
representation of the $SU(N_c)$ gauge group, and $m_f$ are the
quark masses for $f=1,\ldots,N_f$. We focus on the case $N_f<2N_c$
where the $\N=2$ theory is asymptotically free, hence
Seiberg-Witten curves can be used.

In the limit $\mu \ra \infty$, the field $\phi$ decouples and can
be integrated out to leave pure SQCD. Classically, when all $N_f$
quark masses are non-zero, the minima of the SQCD potential is
given by $Q^f=\Qt_f=0$.

The Seiberg-Witten curve for pure $\N=2$ plus fundamental matter
and $SU(N_c)$ gauge group is
\begin{equation}\label{eq:SW_curve}
y^2 = \frac{1}{4}\prod_{a=1}^{N_c} (x-\phi_a)^2 - \La^{2N_c
-N_f}\prod_{f=1}^{N_f}(x+m_f),
\end{equation}
where $\phi=\text{diag }(\phi_1,\ldots \phi_{N_c})$ and
$\sum_{a=1}^{N_c}{\phi_a=0}$.

We can in principle calculate $<Tr \phi^2>$ for the $\N=1$ theory
described by $W_{tree}$ in \eqref{eq:W_tree}, using the
Seiberg-Witten curve. The mass term $\mu\text{Tr }\phi^2$ lifts
the vacuum degeneracy and causes the moduli, $\phi_a$, to move to
the point on the curve where the maximum number of massless
magnetic monopoles condense. These $(N_c-1)$ massless monopoles
are responsible, via the Higgs mechanism, for giving a mass to the
abelian gauge-fields belonging to the $\N=2$ multiplet associated
with $\phi_a$, as argued in \cite{Seiberg:1994aj,Seiberg:1994rs}.

The maximum $(N_c-1)$ massless magnetic monopoles correspond to
$(N_c-1)$ double zeros of the SW curve. So $\phi_a$ are chosen
such that the curve is of the form
\begin{equation}
y^2 = F_2(x) H^2_{N_c-1}(x),
\end{equation}
where the subscript on $F$ and $H$ refers to the degree of the
polynomial in $x$. The tuning of $\phi_a$ such that the curve is
of the above form is known as performing a complete factorization
of the curve.

In practice, the curve with fundamental matter is extremely
difficult to factorize. In fact, it is only recently that the
general solution has been found \cite{Demasure:2002jb}, by
exploiting new powerful tools \cite{Dijkgraaf:2002dh} for
calculating F-terms of the $\N=1$ quantum superpotential. We will
not need these techniques though, as in the limit $\mu \ra
\infty$, the curve \eqref{eq:SW_curve} simplifies to a more
manageable form, where the technique of Chebysev polynomials is
applicable, as in \cite{Konishi:2003ts}.

First we write the $\N=2$ dynamical scale $\La_{{\cal N}=2}$ in
terms of the scale on the curve $\La$
\begin{equation}
\La^2 = 2^{-N_c/(N_c-\half N_f)}\La_{\N=2}^2.
\end{equation}
We then match the dynamical scales of the two theories in the
usual way
\begin{equation}\label{eq:scales_matching}
\La_{\N=1}^{3N_c-N_f} = \mu^{N_c} \La_{\N=2}^{2N_c - N_f}.
\end{equation}

Before taking the large $\mu$ limit, it is convenient to define
new variables
\begin{eqnarray}
\phih_a = \sqrt{\mu} \phi_a, & \xh = \sqrt{\mu} x, & \yh^2 =
\mu^{N_c} y^2.
\end{eqnarray}
In the limit $\mu \ra \infty$, while keeping $\La_{\N=1}$
constant, the Seiberg-Witten curve is given by
\begin{equation}
\yh^2 = \frac{1}{4}\prod_{a=1}^{N_c} (\xh-\phih_a)^2 -
\frac{\La_{\N=1}^{3N_c - N_f}}{2^{N_c}}\det{m},
\end{equation}
where $m=\text{diag }(m_1,\ldots m_{N_f})$, and $\yh^2$ can be
factorized using Chebysev polynomials \cite{Douglas:1995nw}, just
as if there's no fundamental matter.

At the point where there is a maximum number of double zero's, the
curve becomes
\begin{eqnarray}
\yh^2 = \frac{\La_{\N=1}^{3N_c-N_f} \det{m}}{2^{N_c}}
(T_{N_c}^2(\xi) - 1),
\end{eqnarray}
where
\begin{eqnarray}
\xi = (\frac{1}{2^{N_c}\La_{\N=1}^{3N_c-N_f}
\det{m}})^{\frac{1}{2N_c}} \xh e^{-2\pi i k/2N_c} &&
k=1,\ldots,N_c,
\end{eqnarray}
where $T_{N_c}(\xi)$ is a Chebysev polynomial of the first kind,
as described in appendix \ref{sec:chebysev}, and we have used
equation \eqref{eq:T_normalize} to normalize $\xi$. So the
rescaled moduli are
\begin{equation}
\phih_a = \om_a ({2^{N_c}\La_{\N=1}^{3N_c-N_f}
\det{m}})^{\frac{1}{2N_c}} e^{2\pi i k/2N_c},
\end{equation}
where $\om_a$ is defined in equation \eqref{eq:_om_a}. Note that
$\sum_a{\phih_a}=0$, since $\sum_a{\om_a}=0$. Furthermore
\begin{equation}
<\text{Tr } \phih^2> = \lim_{\mu \ra \infty}\mu<\text{Tr } \phi^2>
= N_c (\La_{\N=1}^{3N_c-N_f}\det{m} )^{\frac{1}{N_c}}e^{2\pi i
k/\N_c},
\end{equation}
where we have used that $\sum_{a=1}^{N_c}{\om_a \om_a}=N_c/2$.

Using the Konishi anomaly equations \cite{Cachazo:2003yc}
\begin{eqnarray}
\frac{1}{16 \pi^2}<\text{Tr } \la \la > + <\text{Tr } q^f \phi
\qt_f> &=& \frac{\mu}{N_c} <\text{Tr
}\phi^2>,\\
<\text{Tr } q^f \qt_g> + \sqrt{2} {(m^{-1})^f}_g <\text{Tr }\qt_g
\phi q^f> &=& \frac{{(m^{-1})^f}_g}{16 \pi^2} <\text{Tr }\la \la>,
\end{eqnarray}
where $\la$ is the gaugino field, and $q^f$ and $\qt_g$ the lowest
components of the quark superfields $Q^f$ and $\Qt_g$, we
calculate the gaugino and meson condensates. Note that in the
limit $\mu \ra \infty$, using equation \eqref{eq:scales_matching},
$\La_{\N=2}$ becomes zero and hence so does $\phi_a$. Thus, terms
of the form $<\text{Tr } q \phi \qt>$ can be ignored.

Finally, the gaugino and meson vacuum condensates are given by
\begin{eqnarray}
\frac{1}{16 \pi^2}<\text{Tr } \la \la> =  (\La_{\N=1}^{3N_c - N_f} \det m)^{\frac{1}{N_c}} e^{2\pi i k/\N_c},\\
<\text{Tr } q^f \qt_g> ={(m^{-1})^f}_g (\La_{\N=1}^{3N_c - N_f}
\det m)^{\frac{1}{N_c}} e^{2\pi i k/\N_c},
\end{eqnarray}
respectively, where $k=1 \ldots N_c$.


\begin{acknowledgments}
I would like to thank Paolo Di Vecchia for suggesting the problem,
and Yves Demasure, Paolo Merlatti and Gabriele Ferretti for
helpful discussions. This work was partly funded by a Marie Curie
training site fellowship.
\end{acknowledgments}

\begin{appendix}
\section{The Chebysev polynomials}\label{sec:chebysev}
The Chebysev polynomials  of the first and second kind are
$T_n(\xi)=\cos(n \cos^{-1}{\xi})$ and
$U_n(\xi)=[\sin((n+1)\cos^{-1}{\xi})/\sqrt{1-\xi^2}]$
respectively. They obey the relation
\begin{equation}
T_n^2(\xi)-1 = (\xi^2-1)U_{n-1}^2(\xi)
\end{equation}
The zero's of $T_n$ are given by
\begin{eqnarray}\label{eq:_om_a}
\om_a = \cos{(\frac{\pi}{n}(a-\half))} && a = 1,\ldots,n,
\end{eqnarray}
and of $U_n$ by
\begin{eqnarray}
\zeta_a= \cos{(\frac{\pi}{n+1}a)} && a = 1,\ldots,n,
\end{eqnarray}
Thus,
\begin{eqnarray}
\sum_{a=1}^n{\om_a}=0, && \sum_{a=1}^n{\om_a \om_a}=n/2.
\end{eqnarray}
It is useful to know the explicit factorized form of $T_n$ and
$U_n$
\begin{equation}\label{eq:T_normalize}
T_n(\xi) =
\begin{cases}
\half\prod_{a=1}^{n/2}(4 \xi^2 - 4\om_a^2) & \text{(n
even)}\\
\half(2\xi) \prod_{a=1}^{(n-1)/2}(4 \xi^2 - 4\om_a^2) & \text{(n
odd)},
\end{cases}
\end{equation}
\begin{equation} U_n(\xi) =
\begin{cases}
\prod_{a=1}^{n/2}(4 \xi^2 - 4\zeta_a^2) & \text{(n
even)}\\
(2\xi) \prod_{a=1}^{(n-1)/2}(4 \xi^2 - 4\zeta_a^2) & \text{(n
odd)}.
\end{cases}
\end{equation}

\end{appendix}

\bibliography{condensates}
\end{document}